\documentclass[english,a4paper,aip,jcp,reprint,twocolumn,amsmath,amssymb,superscriptaddress]{revtex4-1}
\pdfoutput=1
\usepackage{epstopdf}
\usepackage{preamble}

\newcommand{\sherwood}{C}

\begin{document}

\title{Nucleation kinetics in drying sodium nitrate aerosols}
\date{\today}

\author{Joshua F. Robinson}
\email{joshua.robinson@bristol.ac.uk}
\affiliation{H.\ H.\ Wills Physics Laboratory, University of Bristol, Bristol BS8 1TL, UK}
\author{Florence K. A. Gregson}
\affiliation{School of Chemistry, Cantock’s Close, University of Bristol, Bristol, BS8 1TS, UK}
\author{Rachael E. H. Miles}
\affiliation{School of Chemistry, Cantock’s Close, University of Bristol, Bristol, BS8 1TS, UK}
\author{Jonathan P. Reid}
\affiliation{School of Chemistry, Cantock’s Close, University of Bristol, Bristol, BS8 1TS, UK}
\author{C. Patrick Royall}
\affiliation{H.\ H.\ Wills Physics Laboratory, University of Bristol, Bristol BS8 1TL, UK}
\affiliation{School of Chemistry, Cantock’s Close, University of Bristol, Bristol, BS8 1TS, UK}
\affiliation{Centre for Nanoscience and Quantum Information, University of Bristol, Bristol BS8 1FD, UK}

\begin{abstract}
  A quantitative understanding of the evaporative drying kinetics and nucleation rates of aqueous based aerosol droplets is important for a wide range of applications, from atmospheric aerosols to industrial processes such as spray drying.
  Here, we introduce a numerical model for interpreting measurements of the evaporation rate and phase change of drying free droplets made using a single particle approach.
  We explore the evaporation of aqueous sodium chloride and sodium nitrate solution droplets.
  Although the chloride salt is observed to reproducibly crystallise at all drying rates, the nitrate salt solution can lose virtually all of its water content without crystallising.
  The latter phenomenon has implications for our understanding of the competition between the drying rate and nucleation kinetics in these two systems.
  The nucleation model is used in combination with the measurements of  crystallisation events to infer nucleation rates at varying equilibrium state points, showing that classical nucleation theory provides a good description of the crystallisation of the chloride salt but not the nitrate salt solution droplets.
  The reasons for this difference are considered. 
\end{abstract}


\maketitle

\section{Introduction}

Understanding the evaporative drying and crystallisation of aerosol droplets is important for a broad range of industrial applications, most notably in spray-drying \cite{OrdoubadiPR2019}, and for predicting the optical and physical properties of atmospheric aerosols \cite{ReidNC2018}.
The microphysics of the drying process also likely has an impact on the viability of bacteria in aerosols \cite{FernandezJRSI2019}.
In spray drying the goal is to control the distribution of sizes, morphology and phase of the final droplets, which are very sensitive to processing conditions such as solvent \cite{CarverIECR2012,LintingreSM2016}, temperature \cite{IveyAST2018,YouDT2014,LinPT2015}, pH \cite{YuJPS2002,DubbiniIJP2014} and additional co-excipients \cite{ZhongAP2018,NandiyantoAPT2011,LyuJCG2017}.
Tailoring crystallisation is particularly important because crystal and amorphous states have fundamentally different properties: crystalline droplets are typically more stable and suitable for product storage \cite{VehringJAS2007,CostantinoJPS1998}, whereas amorphous droplets are more soluble which is desirable for inhalable powders for respiratory drug delivery \cite{AmstadJPCB2016,BroughIJP2013}.
Typically, investigations of crystal nucleation rates can inform the design of spray-drying conditions to achieve a desired final state.

In atmospheric aerosols, the radiative forcing of atmospheric aerosols is strongly influenced by their optical properties \cite{HodasACP2015,CotterellACP2017}.
The solute concentration and physical state (i.e. whether it is crystalline or amorphous) can have an important effect on climate predictions.
In addition, the partitioning of chemical components between the gas and condensed phases is strongly dependent on the phase state of the ambient particles.
This has implications for the long range transport of pollutants, the health impacts of ambient aerosol and the ice nucleation efficiencies of atmospheric particles \cite{ReidNC2018}.

In this work we will investigate drying and crystal nucleation of free aerosol droplets by combining experiments and a numerical model for free aerosol droplets.
The experiments are described in section \ref{sec:experiments}, and we report comparisons with a diffusional model of droplet evolution in section \ref{sec:evolution}.
We find that classical nucleation theory accurately predicts the crystallisation times for \ce{NaCl} aerosols, but not for \ce{NaNO3}, in section \ref{sec:nucleation}.
For \ce{NaNO3} we report nucleation rates with non-monotonic behaviour with increasing solute concentration.

\section{Experiments}
\label{sec:experiments}

The kinetics of drying \ce{NaCl} and \ce{NaNO3} droplets were measured using the Comparative-Kinetics Electrodynamic Balance (CK-EDB).
The CK-EDB instrument has been detailed in previous work \cite{DaviesAST2012} so we will only describe it briefly here.
Droplets of known concentration are produced by a droplet-on-demand generator (MicroFab) and injected into the CK-EDB instrument.
Upon generation the droplets are charged ($<\SI{10}{\femto\coulomb}$ through an ion imbalance) with an induction electrode such that they become trapped within the centre of the electrodynamic field, produced by the application of an AC field between two sets of concentric cylindrical electrodes.
An additional DC field is applied to the lower set of electrodes to counteract gravity and drag forces acting on the droplet.
A circulating current of ethylene glycol coolant across the electrodes controls the chamber temperature $T_{\infty}$ in the range 273--\SI{323}{\kelvin}.

To determine the size and physical state of the droplet, it is illuminated with a \SI{532}{\nano\metre} continuous-wave laser.
The resulting elastic light scattering pattern is recorded by a CCD camera placed at \SI{45}{\degree} to the beam over an angular range of $\SI{\sim24}{\degree}$.
For isotropic droplets in a liquid or dried amorphous state the droplet radius $R$ determines the angular separation between the fringes in the pattern $\Delta \theta$.
Assuming the geometric optics approximation of Mie theory, this relationship is given by
\begin{equation*}
  R
  =
  \frac{\lambda}{\Delta \theta} \left(
  \cos{\left(\frac{\theta}{2}\right)}
  + \frac{n \sin{\left(\frac{\theta}{2}\right)}}{\sqrt{1 + n^2 - 2n \cos{\left(\frac{\theta}{2}\right)}}}
  \right)^{-1}
\end{equation*}
where $\lambda$ is the laser wavelength, $\theta$ is the central viewing angle and $n$ is the droplet refractive index.
This approximation scheme allows estimation of the droplet radius within an accuracy of $\SI{\pm100}{\nano\metre}$.
This method fails when crystallisation occurs breaking isotropy and the scattering pattern dramatically changes; this feature allows the time of crystallisation to be determined to within $\SI{\sim10}{\milli\second}$.
Nucleation and growth occur on such a short time scale that it is not possible to obtain information from the experiments on where inside the droplet this occurs or how many initial nucleation sites there are; we can only determine that the droplet has nucleated crystals.

The instrument features two gas flows for humidified and dry nitrogen applied to the droplet at a rate of \SI{0.03}{\metre\per\second}.
Controlling the ratio of these two flows through a mass-flow controller (MKS instruments) sets the relative humidity (RH) inside the CK-EDB chamber.
Liquid aqueous \ce{NaCl} and aqueous \ce{NaNO3} droplets (20\% solute concentration by weight) were evaporated into dry conditions at \SI{20}{\celsius}.
In all experiments HPLC-grade water, BioXtra $\ge 99.5\%$ \ce{NaCl} (Sigma-Aldrich) and analytic grade \ce{NaNO3} (Fisher-Scientific) were used.
Crystallisation of multiple \ce{NaCl} droplets occurred reproducibly \SI{1}{\second} after droplet generation \cite{GregsonJPCB2019}, whereas \ce{NaNO3} droplets showed stochastic behaviour with a fraction of droplets not crystallising over the timescale of the experiment (droplets were typically trapped for \SI{10}{\second}).
The stochastic behaviour persists when the experiment was repeated for the \emph{same} \ce{NaNO3} droplet over a cycle of repeatedly lowering and raising the RH (described in more detail elsewhere \cite{GregsonTBD2019}), ruling out impurity-driven heterogeneous nucleation.

\begin{figure}[t]
  \includegraphics[width=0.9\linewidth]{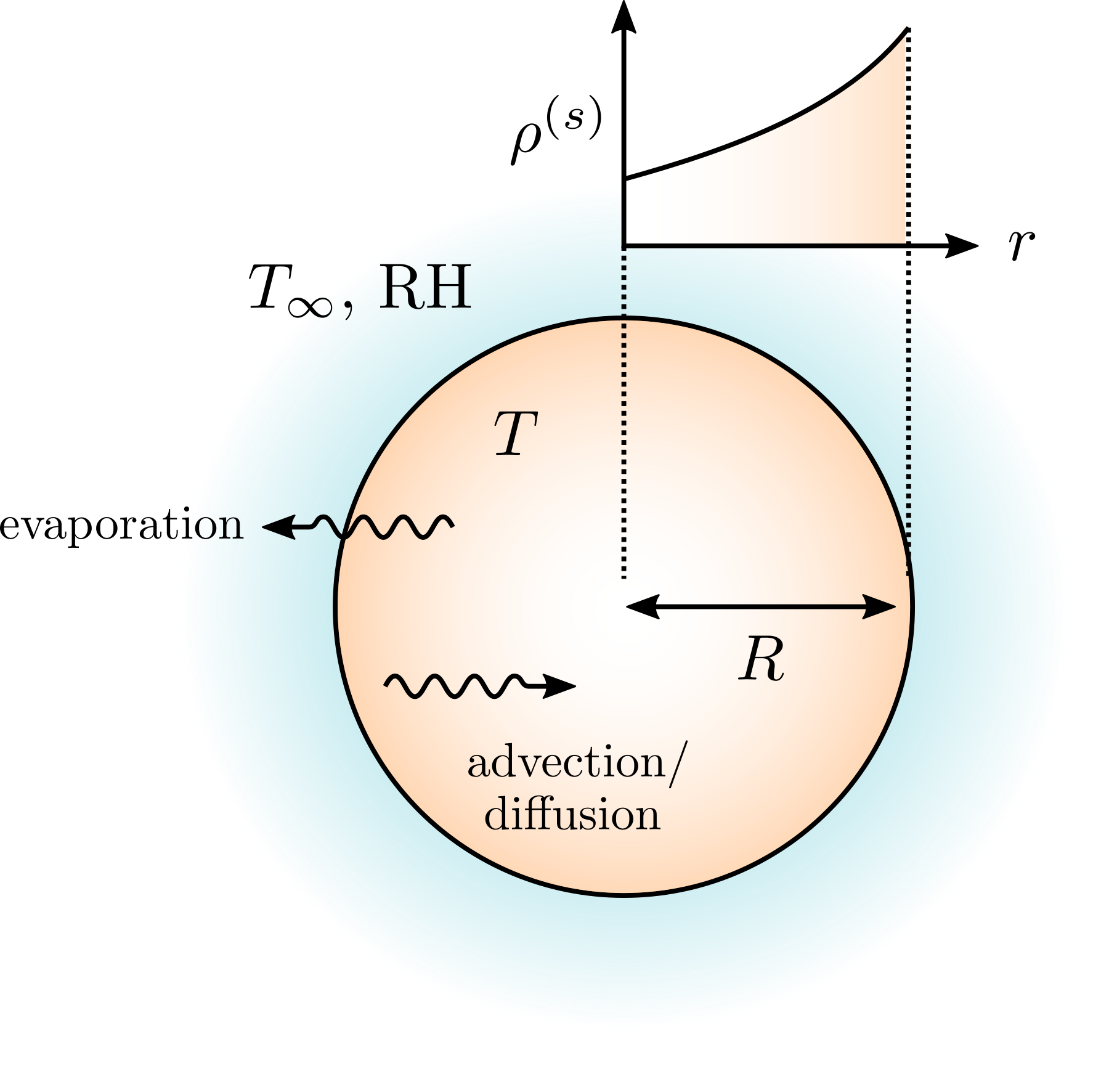}
  \caption{A drying droplet solution of radius $r=R(t)$ surrounded by a gas of temperature $T_\infty$ and relative humidity RH.
    Evaporation of the solvent (water) causes the droplet to shrink and surface enrichment of solute concentration $\rho^{(s)}$ together with evaporative cooling $T < T_\infty$.}
  \label{fig:aerosol-droplet}
\end{figure}

\section{Model for a drying droplet}
\label{sec:evolution}

\subsection{Overview and notation}

In order to obtain nucleation rates we require the time evolution of the droplet's concentration profile over its drying history, and a phenomenological model for nucleation rates based on concentration.
To determine the concentration profile trajectory for a drying droplet we have to consider the relative motion of solute and solvent species inside the droplet, of various species in the surrounding vapour phase, as well as the evaporation of solvent across the phase boundary.
Our approximations will reduce this to a moving boundary problem with solely diffusional mixing.

Prior to crystallisation a drying droplet will be approximately spherical, so we consider a phase boundary at radius $R(t)$ evolving in time $t$.
Writing the distance from the center of the droplet as $r$, the phase boundary separates the liquid phase inside $r \in [0, R(t))$ from the vapour phase outside $r \in (R(t), \infty]$.
The droplet is sketched in Fig.\ \ref{fig:aerosol-droplet}.
In our earlier work \cite{GregsonJPCB2019}, we developed a numerical model for droplet drying which imposes the entire trajectory of $R(t)$ as an input; however, it is difficult to extend this route to long trajectories in a self-consistent way.
Here we improve upon our earlier model by only imposing the initial values of $R$ and its time-derivative, providing an important step towards a unique and entirely first-principles model that both captures the evaporation kinetics (including heat and mass transfer) along with the nucleation kinetics and phase change.

We label the solute and (ambient) gas components as $s$ and $g$ respectively, and the evaporating solvent component as $f$ for \emph{fluid} as it exists in both the liquid and gas phases.
The density is $\rho = \sum_i \rho^{(i)}$ where the (mass) concentration of each component is $\rho^{(i)}$ for $i \in (f,s)$ in the droplet and $i \in (f,g)$ in the gas.
A useful auxillary variable is the mass fraction of each component, i.e.\
\begin{equation}\label{eq:mass-fraction}
  Y^{(i)} = \frac{\rho^{(i)}}{\rho}.
\end{equation}
As the liquid phase is a binary mixture we only need to solve for one component; we choose to solve for the solute mass fraction $Y^{(s)}$.

The thermal conductivity of liquids is generally much larger than the mass diffusivity, so to leading order we can treat temperature $T$ as homogeneous throughout the droplet.
This approximation neglects potential conduction forces driven by temperature gradients.
The droplet temperature will be lower than the ambient temperature $T_\infty$ because vaporisation carries a latent heat, and we determine it self-consistently from the vaporisation rate.
Later we use $T$ in predicting nucleation rates.
However, as a simplification we do not incorporate this temperature into the dynamics themselves through modified diffusion coefficients.
This approximation is reasonable because the fractional temperature change is always less than \SI{5}{\percent}.

\subsection{Evolution of the concentration profile}

In the absence of any chemical reactions the continuity equation for each species component reads
\begin{equation}\label{eq:species-continuity}
  \frac{\partial \rho^{(i)}}{\partial t} +
  \vec{\nabla} \cdot (\rho^{(i)} \vec{v}^{(i)}) = 0
  \qquad i \in \{f,s\}
\end{equation}
where $\vec{v}^{(i)}$ is the velocity of species $i$, or in terms of relative flows
\begin{equation}\label{eq:species-continuity-relative}
  \frac{\partial \rho^{(i)}}{\partial t} +
  \vec{\nabla} \cdot (\rho^{(i)} \vec{v}) +
  \vec{\nabla} \cdot \vec{j}^{(i)} = 0
  \qquad i \in \{f,s\}
\end{equation}
where the mass-averaged fluid velocity is $\vec{v} = \sum_i Y^{(i)} \vec{v}^{(i)}$ and the relative mass flux is $\vec{j}^{(i)} = \rho^{(i)} (\vec{v}^{(i)} - \vec{v})$.
Any advective/convective flows will typically be contained in $\vec{v}$, while diffusive effects are captured by $\vec{j}^{(i)}$.

Volume additivity holds to a good approximation \cite{HandscombCES2009}, i.e.\ the density and concentrations are related by
\begin{equation}\label{eq:volume-additivity}
  \frac{1}{\rho} =
  \frac{Y^{(s)}}{\rho_0^{(s)}} + \frac{Y^{(f)}}{\rho_0^{(f)}},
\end{equation}
where $\rho_0^{(i)}$ is the liquid-phase density of the pure substance; as no stable amorphous phases of \ce{NaCl} or \ce{NaNO3} are known we approximate $\rho_0^{(s)}$ by the density in the crystal phase.

By considering mass conservation one obtains
\begin{equation*}
  \vec{\nabla} \cdot \vec{v}
  =
  \frac{1}{\rho^2}
  \frac{\partial \rho}{\partial Y^{(s)}}
  \vec{\nabla} \cdot \vec{j}^{(s)}
\end{equation*}
so assuming volume additivity \eqref{eq:volume-additivity} we can define the mass difference parameter as
\begin{equation}\label{eq:lambda}
  \Lambda
  =
  \frac{1}{\rho^2} \frac{\partial \rho}{\partial Y^{(s)}} \\
  =
  \frac{1}{\rho^{(f)}_0} -
  \frac{1}{\rho^{(s)}_0}
\end{equation}
giving $\vec{v} = \Lambda \vec{j}^{(s)}$.
This simplifies the advective term in the continuity equation \eqref{eq:species-continuity-relative} leading to
\begin{equation}\label{eq:species-continuity-relative-without-advection}
  \frac{\partial \rho^{(s)}}{\partial t} +
  \vec{\nabla} \cdot \left(
  (1 + \Lambda \rho^{(s)}) \, \vec{j}^{(s)}
  \right) = 0.
\end{equation}
For the relative mass flux we assume Fick's law for diffusion
\begin{equation}\label{eq:ficks-law}
  \vec{j}^{(i)} = -D_\mathrm{eff} \rho \vec{\nabla} Y^{(i)},
\end{equation}
where $D_\mathrm{eff}$ is an effective \emph{binary} diffusion constant for the relative motion.
Inserting \eqref{eq:ficks-law} into \eqref{eq:species-continuity-relative-without-advection} and using the product rule, i.e.\
\begin{equation*}
  \vec{\nabla} \rho^{(s)}
  =
  \left(
  1 +
  \Lambda \rho^{(s)}
  \right)
  \rho
  \vec{\nabla} Y^{(s)}.
\end{equation*}
Finally, this gives the standard diffusion equation
\begin{equation}\label{eq:final-diffusion}
  \frac{\partial \rho^{(s)}}{\partial t}
  =
  \vec{\nabla} \cdot \left(
  D_\mathrm{eff} \vec{\nabla} \rho^{(s)}
  \right),
\end{equation}
where the advective forces have vanished providing a convenient form for numerical implementation.

\begin{figure}[t]
  \includegraphics[width=\linewidth]{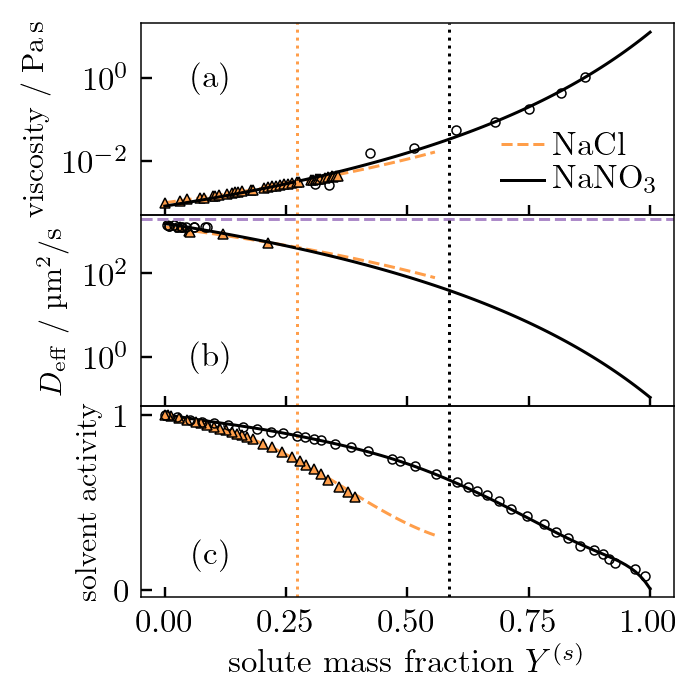}
  \caption{Numerical fits of binary diffusion coefficients for aqueous ionic solutions.
    (a) Fits of viscosity to an Arrhenius-like form \eqref{eq:vft-fit} to experimental values \cite{PowerCS2013,BaldelliAST2016}.
    (b) Diffusion coefficient from the viscosity fits assuming a Stokes-Einstein form \eqref{eq:stokes-einstein}, where the Stokes radius is obtained by calibration with direct measurements of diffusion at \SI{27}{\celsius} for \ce{NaCl} \cite{LyubartsevJPC1996} and \SI{25}{\celsius} for \ce{NaNO3} \cite{YehJCED1970}.
    (c) Solvent activity $a_f$ from a numerical model \cite{CleggJPCA1998} compared with experimental data \cite{TangJGR1994,RovelliJPCA2016}.
    The extrapolations are taken up until the maximum mass fraction explored by the numerical model.
    The dashed-purple horizontal line shows the diffusion constant of pure \ce{H2O} for reference, to be distinguished from the binary diffusion constants in the limit $Y^{(s)} \to 0$.
    The dotted lines indicate the saturation thresholds at \SI{20}{\celsius}.}
  \label{fig:diffusion-fit}
\end{figure}

Bulk viscosity measurements are unavailable for highly concentrated solutions because of the propensity for the salts to crystallise, so we extrapolate the available experimental data \cite{PowerCS2013,BaldelliAST2016} assuming the Arrhenius-like form
\begin{equation}\label{eq:vft-fit}
  \log{\eta}
  =
  \log{\left(\eta(\rho^{(s)} = 0)\right)}
  + \alpha \rho^{(s)},
\end{equation}
where $\alpha$ is a fitting parameter.
The fits are shown in Fig.\ \ref{fig:diffusion-fit}(a).
We model the diffusion constant by assuming the Stokes-Einstein form
\begin{equation}\label{eq:stokes-einstein}
  D_\mathrm{eff} = \frac{k_B T}{6 \pi \eta a},
\end{equation}
where $a$ is the Stokes radius and $\eta$ is the dynamic viscosity.
To determine $a$ we calibrated direct measurements of diffusion from molecular dynamics simulations for \ce{NaCl} \cite{LyubartsevJPC1996} and experiments for \ce{NaNO3} \cite{YehJCED1970} against the viscosity fits.
We obtain $a=\SI{0.169}{\nano\metre}$ for \ce{NaCl} and $a=\SI{0.167}{\nano\metre}$ for \ce{NaNO3}.
The resulting diffusion coefficients entering the droplet evolution equation are shown in Fig.\ \ref{fig:diffusion-fit}(b).

We employ two simplifications in our calculations concerning the effects of droplet temperature.
First, our volume additivity assumption \eqref{eq:volume-additivity} makes the density temperature independent; this neglects conduction forces caused by temperature gradients and results in more heterogeneous droplets.
Secondly, we approximate $T \sim T_\infty$ in the Stokes-Einstein relation \eqref{eq:stokes-einstein}
This approximation neglects evaporative cooling which would slow diffusion, so overestimates the diffusion constant and will result in more heterogeneous droplets.
It is unclear \emph{a priori} which of these opposing effects dominates.
Note: later we model the droplet temperature $T$ explicitly for treating solvent evaporation and nucleation rates, but we have not incorporated this temperature into the diffusion constant.

\subsection{Droplet boundary conditions}

Initially, the droplets are prepared as equilibrium solutions, so they are well-mixed and we can assume a uniform initial concentration profile.
At $t=0$ a droplet is produced which begins to lose solvent through evaporation due to the low RH of the CK-EDB.
The evaporation rate determines the boundary conditions for the diffusion equation \eqref{eq:final-diffusion}.

Integrating the species continuity equation \eqref{eq:species-continuity} gives the total mass flow into the droplet (of each species) as
\begin{equation}\label{eq:integral-continuity}
  \frac{d m^{(i)}}{dt} =
  \int_{V(t)} \frac{\partial \rho^{(i)}}{\partial t} \, d V +
  \int_{\partial V(t)} \rho^{(i)} \, \vec{v}_{\partial V(t)} \cdot d\vec{S},
\end{equation}
where $V(t)$ is the volume of the droplet at time $t$ and $\vec{v}_{\partial V(t)}$ is the velocity of the boundary, and the vectorial surface element $d\vec{S}$ points in the direction of the outer normal vector.
We assume the solute does not leave the droplet, so all mass flow at the boundary must be due to the solvent.
Inserting the diffusion equation \eqref{eq:final-diffusion} into \eqref{eq:integral-continuity} and applying Stokes' theorem gives
\begin{equation}
  \frac{d m^{(s)}}{dt}
  =
  \int_{\partial V(t)}
  \Big(
  \rho^{(i)} \vec{v}_{\partial V(t)} +
  D_\mathrm{eff} \vec{\nabla} \rho^{(s)}
  \Big)
  \cdot d\vec{S}
  = 0.
\end{equation}
For spherical droplets this gives the boundary condition
\begin{equation}
  \left. \frac{\partial \rho^{(s)}}{\partial r} \right|_{r=R(t)}
  =
  -
  \frac{\rho^{(s)}}{D_\mathrm{eff}(R)}
  \frac{dR}{dt}.
\end{equation}
Assuming volume additivity \eqref{eq:volume-additivity} we can determine the radial evolution from mass conservation as
\begin{equation}\label{eq:radial-evolution}
  \frac{dR}{dt}
  =
  \frac{1}{4\pi R^2 \rho^{(f)}_0} \frac{dm^{(f)}}{dt}
\end{equation}
so we need a model for the evaporation rate $\tfrac{dm^{(f)}}{dt}$ to close this system of equations.

We assume the classical result for quasistatic vaporisation \cite{Slattery1999,Sirignano2010}
\begin{equation}\label{eq:quasistatic-vaporisation}
  \frac{dm^{(f)}}{dt}
  =
  4\pi \rho_v D_v R \ln{(1 + B)}.
\end{equation}
where $\rho_v$ and $D_v$ are the density and diffusion constant in the vapour phase, and the \emph{Spalding number} is defined as
\begin{equation}\label{eq:spalding-number}
  B
  =
  \frac{
    \displaystyle{\lim_{r \to \infty}} Y^{(f)}(r) - Y^{(f)}(R^+)
  }{
    1 - \displaystyle{\lim_{r \to \infty}} Y^{(f)}(r)
  }.
\end{equation}
with $Y^{(f)}(R^+)$ indicating that it is the mass fraction of the solvent component on the vapour side of the boundary.
For the phase boundary it is convenient to work with mole fraction instead of $Y^{(f)}(R^+)$, because it can be related to partial pressure $p_f$ through Dalton's law for ideal gases, i.e.\
\begin{equation*}
  X^{(f)}(R^+) = \frac{p_f(R^+)}{p},
\end{equation*}
with $p$ as the total pressure.
This can be converted back into mass fraction for use in \eqref{eq:spalding-number} through
\begin{equation}\label{eq:mole-fraction-conversion}
  Y^{(f)} = \frac{M_f X^{(f)}}{M_f X^{(f)} + M_g (1 - X^{(f)})},
\end{equation}
where $M_i$ are the molar masses of each species.
We can obtain the partial pressure from the solvent concentration at the boundary from the solvent activity, defined through
\begin{equation*}
  a_f(R^-)
  :=
  a_f\Big( Y^{(f)}(R^-) \Big)
  =
  \frac{p_f(R^+)}{p_\mathrm{eq}^*(T)}
  =
  \frac{p \, X^{(f)}(R^+)}{p_\mathrm{eq}^*(T)},
\end{equation*}
where $p_\mathrm{eq}^*$ is the equilibrium vapour pressure of the evaporating component.
Fig.\ \ref{fig:diffusion-fit}(c) shows $a_f$ as a function of mass fraction, obtained through a numerical method that treats the non-ideality of the solution \cite{CleggJPCA1998}.
The Clausius-Clapeyron relation connects the vapour pressure at the surface to the ambient conditions via
\begin{equation*}
  p_\mathrm{eq}^*(T)
  =
  p_\mathrm{eq}^*(T_\infty)
  \exp{\left( \frac{L}{R_g} \frac{T - T_\infty}{T T_\infty} \right)},
\end{equation*}
where $L$ is the specific latent heat of vaporisation and $R_g$ is the molar gas constant.
Combining the above expressions gives the mole fraction above the surface as
\begin{equation}\label{eq:molar-fraction}
  X^{(f)}(R^+) =
  \frac{a_f(R^-) \, p_\mathrm{eq}^*(T_\infty)}{p}
  \exp{\left( \frac{L}{R_g} \frac{T - T_\infty}{T T_\infty} \right)},
\end{equation}
which requires an equation for droplet temperature $T$ for closure.
As a simplification, we ignored curvature effects on $p_\mathrm{eq}^*(T)$ emerging from surface tension \cite{ThomsonPRSE1872}.

Finally, assuming a steady state heat flux through the boundary, and neglecting the radiative heat transfer and the droplet heat capacity, gives the temperature difference between the droplet surface and the ambient temperature as \cite{KulmalaJAS1992,RovelliJPCA2016}
\begin{equation}\label{eq:temperature-difference}
  T - T_\infty = \frac{L}{4\pi R K} \frac{dm^{(f)}}{dt},
\end{equation}
where $K$ is the thermal conductivity of the vapour phase, closing the equations at the phase boundary.
Together, Eqs.\ \eqref{eq:quasistatic-vaporisation}, \eqref{eq:spalding-number}, \eqref{eq:mole-fraction-conversion}, \eqref{eq:molar-fraction} and \eqref{eq:temperature-difference} form a complete set of equations that can be solved (numerically) to obtain the evaporation rate.

Typically the classical vaporisation rate equation \eqref{eq:quasistatic-vaporisation} requires semi-empirical corrections to treat more complex mass and heat transport phenomena at the boundary.
In order to better match the experiments, we introduce the empirical factor $\sherwood$ to correct the vaporisation rate giving
\begin{equation}\label{eq:sherwood-correction}
  \frac{dm^{(f)}}{dt}
  =
  4\pi \, \sherwood \, \rho_v D_v R \ln{(1 + B)}.
\end{equation}
We determine $\sherwood$ from the initial value of $\tfrac{dR}{dt}$ in the experiments.
At constant vaporisation rate the solution to the radial evolution equation \eqref{eq:radial-evolution} yields
\begin{equation}
  R(t)^2
  =
  R(t=0)^2
  + \left(2R \, \frac{dR}{dt}\right)_{t=0} t,
\end{equation}
valid at short times.
We iteratively solve Eqs.\
\eqref{eq:radial-evolution}, \eqref{eq:spalding-number}, \eqref{eq:mole-fraction-conversion}, \eqref{eq:molar-fraction}, \eqref{eq:temperature-difference} and \eqref{eq:sherwood-correction} with varying $\sherwood$ until a value is obtained which produces a $\tfrac{dR}{dt}$ consistent with the experimental fit.
We give the explicit values of $\sherwood$ in the appendix.

\subsection{Implementation and results}

\begin{figure}
  \includegraphics[width=\linewidth]{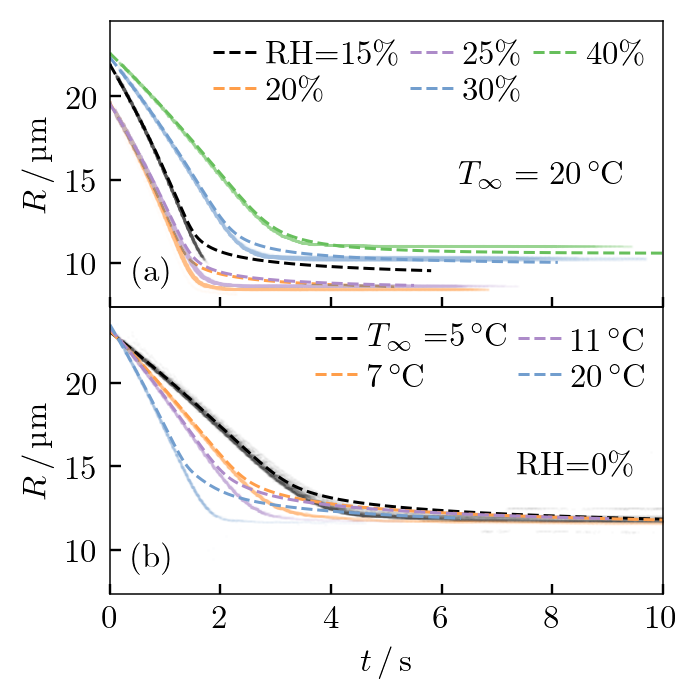}
  \caption{Evolution of \ce{NaNO3} aerosol droplet radii from the numerical model (dashed lines) and experiments shown by points with 1\% transparency, showing reasonable agreement at short times until longer times when the evaporation rate is underestimated.
    (a) Varying relative humidity while ambient temperature is kept fixed, for an initial solute mass fraction of $Y^{(s)} = 0.125$.
    (b) Varying ambient temperature while relative humidity is kept fixed, for an initial solute mass fraction of $Y^{(s)} = 0.2$.}
  \label{fig:nano3-trajectory}
\end{figure}

\begin{figure}
  \includegraphics[width=0.9\linewidth]{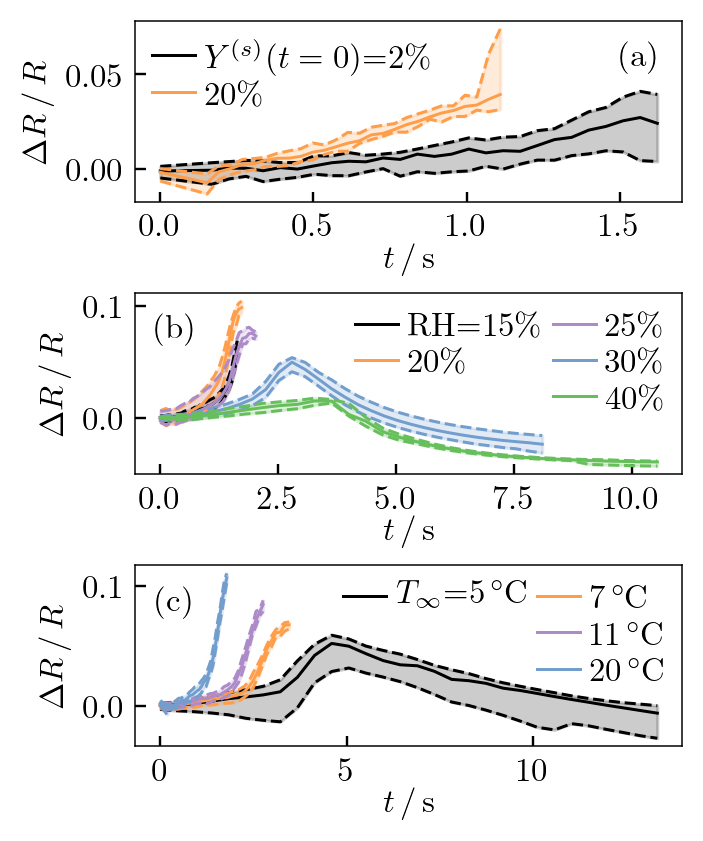}
  \caption{Time evolution of simulated droplet radius error by comparison with experiments.
    We collate data from multiple experiments at each state point; the median across the datasets (solid lines) is shown along with a shaded region indicating agreement up to the 10/90th percentiles (dashed lines).
    (a) \ce{NaCl} solution droplets with ambient temperature $T_\infty = \SI{20}{\celsius}$ in dry conditions.
    (b) \ce{NaNO3} solution droplets with initial solute mass fraction of $Y^{(s)} = 0.125$ and ambient temperature $T_\infty = \SI{20}{\celsius}$.
    (c) \ce{NaNO3} solution droplets with initial solute mass fraction of $Y^{(s)} = 0.2$ in dry conditions.
  }
  \label{fig:radius-errors}
\end{figure}

We discretise the solute concentration profile $\rho^{(s)}(r)$ onto a uniformly spaced grid over $r \in [0, R(t)]$.
To handle the moving boundary it is convenient to work in the rescaled coordinate $\widetilde{r} = \frac{r}{R(t)} \in [0, 1]$.
For the discretisation we define the vector $\vec{\rho} := \{\rho_0, \rho_1, \cdots, \rho_N\}$ where $\rho_i := \rho^{(s)}\left(\widetilde{r} = \frac{i}{N}\right)$.
The complete history of the evolution of the droplet then involves both $\vec{\rho}$ and $R$ variables.
In addition, it is convenient to introduce $\dot{R}$ as its own variable so that the final Jacobian for the diffusion equation \eqref{eq:final-diffusion} has tridiagonal form.
This gives us the evolving droplet state variable $\vec{x} = (\vec{\rho}, R, \dot{R})$.

To integrate a timestep $\Delta t$ we use the Crank-Nicolson \cite{CrankMPCPS1947} method where
\begin{equation*}
  \frac{\vec{x}(t + \Delta t) - \vec{x}(t)}{\Delta t}
  =
  \frac{1}{2}
  \left(
  \left. \frac{\partial \vec{x}}{\partial t} \right|_{t + \Delta t}
  +
  \left. \frac{\partial \vec{x}}{\partial t} \right|_t
  \right)
  + \mathcal{O}(\Delta t^2).
\end{equation*}
As the evolution equations are nonlinear this must be solved iteratively to find a self-consistent solution.
Introducing the $k$th approximation for $\vec{x}(t + \Delta t)$ as $\vec{x}^{(k)}(t + \Delta t)$, we write the next term in the sequence as $\vec{x}^{(k+1)} = \vec{x}^{(k)} + \delta \vec{x}^{(k)}$ and we obtain
\begin{equation*}
  \frac{\vec{x}_{n+1}^{(k)} + \delta\vec{x}_{n+1}^{(k)} - \vec{x}_n}{\Delta t}
  =
  \frac{1}{2}
  \left(
  \frac{\partial (\vec{x}_{n+1}^{(k)} + \delta\vec{x}_{n+1}^{(k)})}{\partial t}
  +
  \frac{\partial \vec{x}_n}{\partial t}
  \right),
\end{equation*}
using the subscript $n$ as shorthand for the time.
This is a matrix equation that can be inverted for $\delta \vec{x}^{(k)}$.
Convergence is deemed to occur where $\delta \vec{x}^{(k)}$ falls below some threshold value.
The main advantage of this scheme over more simple schemes (e.g. forward Euler method where just the initial $\frac{\partial \vec{x}_n}{\partial t}$ is taken) is that the error is of order $\Delta t^2$ ensuring rapid convergence with small timesteps.

We integrated initially homogeneous droplets of \ce{NaCl} and \ce{NaNO3} for various ambient conditions.
The resulting radius, illustrated for \ce{NaNO3} in Fig.\ \ref{fig:nano3-trajectory}; we see that at short times there is excellent agreement because of the introduction of the correcting factor $\sherwood$ in \eqref{eq:sherwood-correction}.
However, at longer times the evaporation rate underestimated.
This is likely due to limitations of the simplified evaporation model \eqref{eq:sherwood-correction} or because the neglect of conductive forces causes the evaporation to become diffusion-limited at long-times when the surface is highly enriched.
We achieve good agreement with experiments for \ce{NaCl} across their entire time evolution (Fig.\ \ref{fig:nacl-trajectory}(b)) because these droplets crystallise before the slowdown of the evaporation rate.
To make this analysis more quantitative we show the errors in the droplet radius in Fig.\ \ref{fig:radius-errors}; we find that the error is always within 10\% in our model throughout the evolution.

\section{Nucleation model}
\label{sec:nucleation}

\subsection{Droplet nucleation rates}

\begin{figure}
  \includegraphics[width=\linewidth]{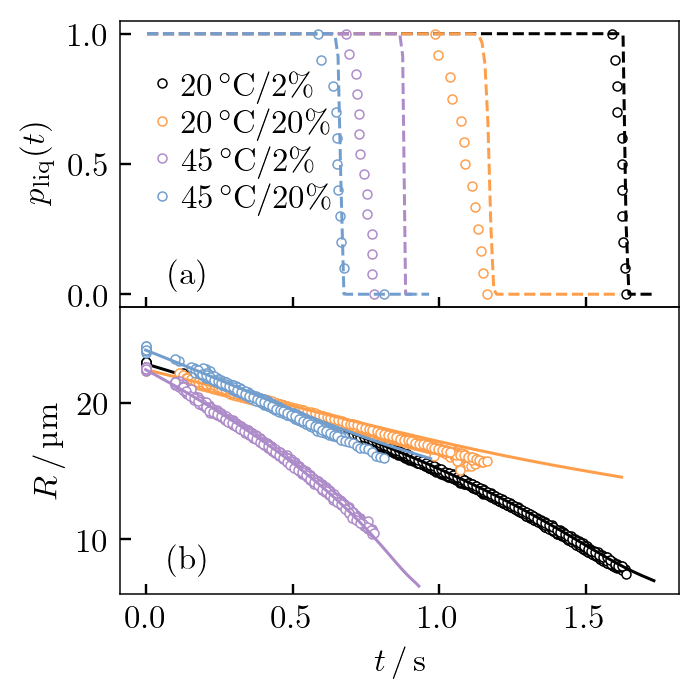}
  \caption{Evolution of \ce{NaCl} droplets in dry air $\mathrm{RH}=0\%$ from experiments (points) and the numerical model (lines) for different ambient temperatures and initial solute mass fractions.
    (a) Probability that a droplet survives without nucleating, assuming the liquid-crystal surface tension $\gamma = \SI{0.08}{\newton\per\metre}$ \cite{DesarnaudJPCL2014} for the numerical model.
    (b) Evolution of droplet radius showing good treatment of solvent evaporation rates.
  }
  \label{fig:nacl-trajectory}
\end{figure}

Denoting the rate of solute nucleation per unit volume as $J$, the continuum limit nucleation rate for the entire droplet is
\begin{equation}\label{eq:droplet-nucleation-rate}
  W
  =
  \int_V J dV
  =
  4\pi \int_0^R J(r) \, r^2 dr.
\end{equation}
Both the local $J$ and total rates $W$ contain an implciit time dependence because of their dependency on the evolving variables $R$, $\rho^{(s)}$ and $T$.
For homogeneous nucleation $J$ depends solely on the state variables $\rho^{(s)}$ and $T$.
Nucleation rates are typically strongly concentration dependent \cite{ValerianiJCP2005,DesarnaudJPCL2014,SearJPCM2007}, so we anticipate nucleation to occur at the boundary $r=R(t)$ where the solute concentration is greatest.
Allowing for heterogeneous nucleation $J$ could acquire an additional dependence on the inhomogeneities in the system; as the experiments were performed with high-purity precursor compounds to mitigate the effect of chemical impurities, we expect the main potential site for heterogeneous nucleation to be the liquid-air interface.
Whichever nucleation mechanism dominates, we expect it to occur at the boundary so the total rate \eqref{eq:droplet-nucleation-rate} reduces to
\begin{equation}\label{eq:boundary-nucleation-rate}
  W
  \sim
  4\pi R^2 J \xi,
\end{equation}
where $J$ is now evaluated at the boundary, and we introduced $\xi$ as the thickness of the typical shell region over which nucleation occurs.
We will give nucleation rates in terms of $J \xi$, assuming a value $\xi = \SI{1}{\micro\metre}$ to set the absolute scale of the rates predicted by theory (section \ref{sec:cnt}) to most closely match the experiments.

We can relate the nucleation rates to the experimentally observed events by assuming Poisson statistics.
We define the survival probability as
\begin{equation*}
  p_\mathrm{liq}(t)
  :=
  \textrm{Prob}\left[ \textrm{no nucleation by time } t \right],
\end{equation*}
The mean number of nucleation events in the time interval $\Delta t$ is simply $W \Delta t$, giving the probability that there is no nucleation event after a time $\Delta t$ as
\begin{equation*}
  p_\mathrm{liq}(t + \Delta t) = p_\mathrm{liq}(t) e^{-W \Delta t}.
\end{equation*}
Taking the infinitesimal limit and using the fact that droplets are prepared in the liquid state giving the initial condition $p_\mathrm{liq}(t=0) = 1$ yields
\begin{equation}\label{eq:survival}
  p_\mathrm{liq}(t)
  =
  \exp{\left( -\int_0^t W \, dt \right)}.
\end{equation}
As we have already determined the droplet's radius and concentration profile from the evolution equations described in section\ \ref{sec:evolution}, we are left needing a model for the nucleation rate per unit volume $J$ before we can determine $p_\mathrm{liq}$.

\subsection{Nucleation models}
\label{sec:cnt}

For nucleation processes with a single barrier the rate per unit volume goes as
\begin{equation}\label{eq:nucleation-rate-barrier}
  J = \kappa \exp{\left(-\frac{\Delta G^{*}}{k_B T}\right)},
\end{equation}
where $\kappa$ is a kinetic prefactor and $\Delta G^*$ is thermodynamic barrier for the process.
A widely used approximation for the kinetic prefactor is \cite{SearJPCM2007}:
\begin{equation}
  \kappa = n_I j Z,
\end{equation}
$n_I$ is the number density of potential nucleation sites, $j$ is the rate of aggregation to these sites, and $Z$ is the Zeldovich factor.
These last two quantities are typically further approximated as \cite{SearJPCM2007}
\begin{subequations}
  \begin{align}
    j &\sim n D_\mathrm{eff} R^*,
    \\
    Z &\sim (N^*)^{-\tfrac{2}{3}},
  \end{align}
\end{subequations}
where $n$ is the solute number density, $N^*$ is excess number of molecules in the critical nucleus and $R^*$ is its radius.
The barrier $\Delta G^*$ depends on the specific nucleation mechanism.

For homogeneous nucleation the sites of nucleation are simply the solute molecules themselves so $n_I = n$.
The driving force for the transition is the chemical potential change $\Delta \mu$ from formation of the new phase.
In classical nucleation theory (CNT) the surface tension between the crystal and liquid is imagined as the main obstacle to nucleation.
Combining the two contributions leads to the barrier
\begin{equation}
  \Delta G = \gamma A - |\Delta \mu| n_c V,
\end{equation}
where $\gamma$ is the liquid-crystal surface tension, $n_c$ is the crystal number density, and $A, V$ are the surface areas and volumes of the nucleated region.
The thermodynamic barrier to nucleation is then the maximum of this formula; assuming a perfectly spherical crystal seed this gives
\begin{align}\label{eq:cnt-barrier}
  \Delta G^{*} &= \frac{4}{3} \pi (R^{*})^2 \gamma, \\
  (R^{*}) &= \frac{2\gamma}{n_c |\Delta\mu|}.
\end{align}
The chemical potential expressed in terms of mean ionic activity is\cite{DesarnaudJPCL2014}
\begin{equation}
  \frac{\Delta \mu}{k_B T}
  =
  2 \ln{\left( \frac{a_\pm}{a_{0\pm}} \frac{\rho^{(s)}}{\rho^{(s)}_0} \right)},
\end{equation}
where $a_\pm$ is mean ionic activity coefficient, $a_0$ is its value at saturation and $\rho^{(s)}_0$ is the threshold saturation concentration.

\begin{figure}
  \includegraphics[width=\linewidth]{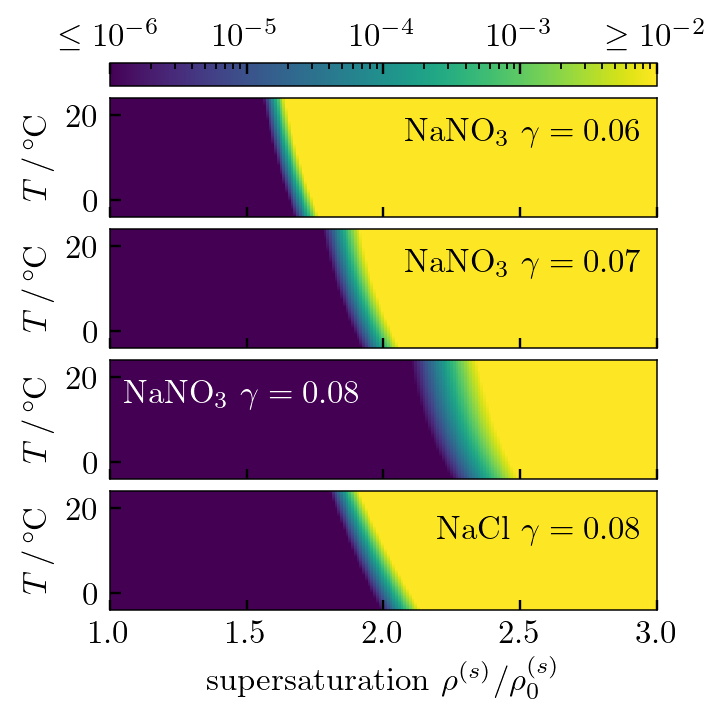}
  \caption{Shell nucleation rate $J\xi$ (\si{\per\micro\metre\squared\per\second}) predicted by classical nucleation theory for aqueous \ce{NaNO3} and \ce{NaCl} solutions at different state points.
    The dark blue and bright yellow regions show where nucleation is essentially impossible or instantaneous on the experimental timescale.
    Different values of solid-liquid surface tension $\gamma$ (given in \si{\newton\per\metre}) do not result in a different qualitative picture: a nucleation rate which monotonically increases with supersaturation and temperature.}
  \label{fig:cnt}
\end{figure}

We find that CNT predicts homogeneous nucleation rates which increase monotonically in both concentration and temperature.
In Fig.\ \ref{fig:cnt} we show the predicted rates for \ce{NaCl} with $\gamma=\SI{0.08}{\newton\per\metre}$ from the literature \cite{DesarnaudJPCL2014} and \ce{NaNO3} with different trial values of surface tension to test correspondence with the experimental data; we find that the nucleation rates are essentially described by a step function of infinite magnitude over the timescale of the experiments.
This is consistent with observations for \ce{NaCl}, so we are able to accurately predict the time of nucleation in the experiments shown in Fig.\ \ref{fig:nacl-trajectory}(a).
By contrast, the experiments show that the final survival probability for \ce{NaNO3} droplets is often in the range $0 < p_\mathrm{liq} < 1$ which is not consistent with nucleation rates being characterised by a step function, which we will make more quantitative in the next section.

\begin{figure}
  \includegraphics[width=\linewidth]{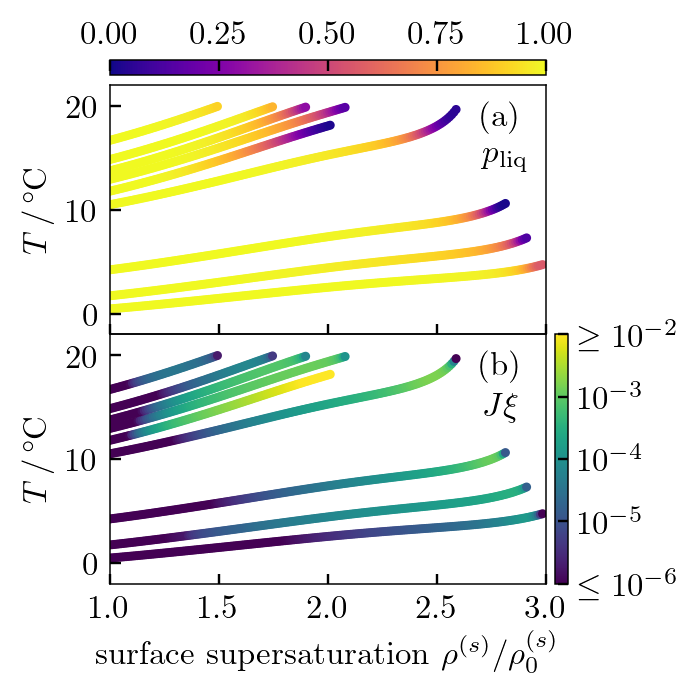}
  \caption{State points explored by experiments with drying \ce{NaNO3}--\ce{H2O} aerosol droplets as determined from our numerical model for 9 datasets for droplet evaporation under different initial conditions.
    (a) Survival probability in the experimental trajectories (i.e.\ the probability that a droplet has not crystallised), with state-point inferred from the model.
    (b) Shell nucleation rates $J\xi$ (\si{\per\micro\metre\squared\per\second}) inferred from trajectories assuming boundary-dominated nucleation \eqref{eq:boundary-nucleation-rate}, showing non-monotonic behaviour in increased concentration and temperature in contrast with the predictions of classical nucleation theory in Fig.\ \ref{fig:cnt}.
  }
  \label{fig:state-points}
\end{figure}

\subsection{Inferring nucleation rates from experiments}

We can try to determine the nucleation rates directly from experiments by observing the stochastic nucleation behaviour over repeat trajectories and comparing these against the numerical model.
The experiments give us the true survival probabilities $p_\mathrm{liq}$ of which we can determine the droplet nucleation rate $W$ exactly by numerical differentiation.
Combined with the numerical model, which gives us the precise state of the droplet, we can infer $J\xi$ from inversion of the rate formula \eqref{eq:boundary-nucleation-rate} under the assumption that nucleation is boundary-dominated.

Differentiation of the survival probability \eqref{eq:survival} yields
\begin{equation}
  \dot{p}_\mathrm{liq}
  =
  - W p_\mathrm{liq},
\end{equation}
upon combining this with our assumption that nucleation occurs near the boundary \eqref{eq:boundary-nucleation-rate} allows to write the nucleation rate as
\begin{equation}
  J\xi
  =
  - \frac{1}{4\pi R^2} \frac{\dot{p}_\mathrm{liq}}{p_\mathrm{liq}},
\end{equation}
which we can determine from the experimentally observed $p_\mathrm{liq}$ trajectory.
The derivative of $p_\mathrm{liq}$ can be obtained through fitting.
The survival probabilities decay monotonically as a generalised step function, so we fit the experimental trajectories with the Fermi-Dirac form
\begin{equation}
  p_\mathrm{liq}(t) - \lim_{t \to \infty} p_\mathrm{liq}(t)
  =
  \frac{1 - \lim_{t \to \infty} p_\mathrm{liq}(t)}{\exp{\left[\epsilon(t - t_s)\right]} + 1},
\end{equation}
where $t_s$ is the time at which saturation is reached $\rho^{(s)}(R) = \rho^{(s)}_0$, and introducing the fitting function
\begin{equation*}
  \epsilon(t)
  =
  \begin{cases}
    at + bt^2 - c / t & \; t > 0 \\
    - \infty & \; t < 0
  \end{cases}
\end{equation*}
subject to the constraint that the fitting parameters $a, b, c \in [0, \infty]$ to ensure that $p_\mathrm{liq}$ decreases monotonically from $p_\mathrm{liq}(t=0)=1$.

In Fig.\ \ref{fig:state-points}(a) we show the survival probabilities for the experiments with \ce{NaNO3} droplets, and we perform the inversion procedure described above to infer bulk nucleation rates in Fig.\ \ref{fig:state-points}(b).
The resulting nucleation rates show non-mononic behaviour, increasing to a maximum before decreasing to essentially zero over the duration of the experiment.
This results in a finite final survival probability $p_\mathrm{liq} > 0$, and starkly contrasts with the picture captured by CNT and realised in \ce{NaCl} droplets (Fig.\ \ref{fig:nacl-trajectory}(a)) where $p_\mathrm{liq}$ would remain close to unity for most of the experiment before sharply dropping to zero as all the droplets crystallise reproducibly.
Fig.\ \ref{fig:cnt} and Fig.\ \ref{fig:state-points}(b) are shown with identical ranges to highlight this contrasting behaviour.

Clearly the nucleation kinetics in drying \ce{NaNO3} aerosols are more complicated than the simple homogeneous nucleation scenario we assumed in section \ref{sec:cnt}.
One kinetic effect we have poorly estimated is the slowing down of diffusion occurring at very high concentrations.
We have assumed the Stokes-Einstein relation holds in this highly saturated regime, which may not be a valid assumption; however, more accurate knowledge of the diffusion constant would only shift the nucleation rates by an order of magnitude, which is insignificant compared to the dramatic (and monotonic) kinetic changes emerging from CNT as seen in Fig.\ \ref{fig:cnt}.
For this reason nucleation in drying $\ce{NaNO3}$ aerosols must occur through a qualitatively different kinetics.
More exotic nucleation processes involve e.g.\ more sophisticated core geometries or pathways featuring multiple steps \cite{SearJPCM2007}.
Such processes may involve multiple reaction coordinates, whereas classical nucleation theory has a single one.

\section{Conclusions}

We have developed a numerical model based on a diffusion equation with an extrapolation of the diffusion constant to high concentrations assuming the Stokes-Einstein relation.
As input the model takes only the initial droplet state, and the resulting evolution conforms well to the experimental trajectories.
Assuming boundary dominated nucleation we are able to predict nucleation rates inside the droplet from CNT, and by inverting this process we can infer the actual observed nucleation rates at varying state points.
The nucleation rates are highly dependent on the rate of droplet drying, as this determines the state points which are ultimately explored.

We found that CNT works well for predicting crystal nucleation in \ce{NaCl} but not \ce{NaNO3} aerosols.
In both cases CNT predicts nucleation essentially after a threshold surface saturation is reached, whereas experiments show nucleation in \ce{NaNO3} has stochastic behaviour.
This emerges from the fact that nucleation rates predicted by CNT monotonically increase in concentration and temperature.
In particular, the change in nucleation rate from increased concentration is so dramatic that the behaviour of CNT is essentially unchanged by small adjustments to the model parameters.

CNT is a model for homogeneous nucleation, so it is possible that it fails because crystallisation occurs for \ce{NaNO3} through heterogeneous nucleation.
The same stochastic phenomena are observed when repeating the experiments with the same droplet on a cycle of decreasing and increasing the RH to dry and then re-condense the droplet; this rules out heterogeneous nucleation through impurities, as the chemical makeup is the same in each cycle yet the phenomenon persists.
This leaves the gas-liquid phase boundary itself as a site for hetereogeneous nucleation.

It is highly likely that the model overestimates the surface enrichment because at long times the simulated evaporation rates become limited by solute diffusion at the boundary.
The diffusion limit would persist even if more sophisticated transport phenomena were introduced to the evaporation model.
Surface enrichment is overestimated because we have neglected the effect of temperature gradients inside the droplet, and because we have used an extrapolation of low concentration diffusion data which likely to underestimate diffusion at high concentrations.
Temperature gradients create inward convection currents reducing surface enrichment.
Our extrapolation of the diffusion constant assumed the Stokes-Einstein relation holds across the whole state space, however this relation can be violated at high viscosities \cite{BerthierRMP2011}.
The rapid increase of viscosity with salt concentration in our model leads to a feedback loop where diffusion becomes increasingly difficult as the surface is enriched.
Correcting for these effects, we expect the surface concentrations explored by the experiments to increase to a maximum before decreasing which could explain non-monoticity.
However, this can only partially explain the observed behaviour because CNT is extremely sensitive to concentration.
Fundamentally, we require a deeper understanding of the nucleation kinetics at ultrahigh supersaturations in order to correctly model the crystallisation of droplet drying.

This work is important in showing that the nucleation rate of nitrate aerosol is not only influenced by the level of supersaturation, but also by the drying kinetics itself because of an interplay between the inhomogeneity of the concentration profile and droplet temperature.
This is important for climate predictions where an understanding of the phase of atmospheric aerosol is crucial, and also valuable for spray-drying models where control over the resulting phase could be enabled by tuning the various drying parameters.

\appendix*

\section{Explicit values of empirical correction}

\begin{ruledtabular}
  \begin{tabular}{rrrr}
    $Y^{(s)}(t=0)$ & T (\si{\celsius}) & RH (\%) & $\sherwood$ \\
    \hline
    \multicolumn{4}{c}{\ce{NaCl} aerosol} \\
    \hline
    0.02 & 20 & 0.00 & 0.762 \\
    0.02 & 45 & 0.00 & 0.463 \\
    0.20 & 45 & 0.00 & 0.326 \\
    0.20 & 20 & 0.00 & 0.608 \\
    \hline
    \multicolumn{4}{c}{\ce{NaNO3} aerosol} \\
    \hline
    0.125 & 20 & 0.15 & 0.825 \\
    0.125 & 20 & 0.20 & 0.780 \\
    0.125 & 20 & 0.25 & 0.793 \\
    0.125 & 20 & 0.30 & 0.735 \\
    0.125 & 20 & 0.40 & 0.654 \\
    0.200 & 20 & 0.00 & 0.651 \\
    0.200 & 10.8 & 0.00 & 0.693 \\
    0.200 & 4.8 & 0.00 & 0.603 \\
    0.200 & 7.4 & 0.00 & 0.714 \\
    0.011 & 11.45 & 0.00 & 0.792 \\
    0.019 & 11.45 & 0.00 & 0.824 \\
    0.052 & 11.45 & 0.00 & 0.854 \\
    0.200 & 11.45 & 0.00 & 0.853 \\
    0.350 & 11.45 & 0.00 & 0.820
  \end{tabular}
\end{ruledtabular}

\,

\begin{acknowledgments}
  FKAG, REHM and JPR acknowledge support from the Engineering and Physical Sciences Research Council under grant code EP/N025245/1.
  JFR and CPR  acknowledge the European Research Council under the FP7 / ERC Grant Agreement No.\ 617266 ``NANOPRS''.
\end{acknowledgments}

\bibliography{bibliography,unpublished}

\end{document}